\documentclass[conference]{IEEEtran}

\usepackage{definitions}
\usepackage{fst}

\newcommand{\Pavg}{P_\text{avg}}
\newcommand{\n}{N}
\newcommand{\kc}{K}
\newcommand{\Ical}{\mathcal{I}}
\newcommand{\Jfun}{\operatorname{J}}
\newcommand{\FER}{\operatorname{FER}}

\begin{document}

\title{Design of Polar Codes for Parallel Channels with an Average Power Constraint}

\author{\IEEEauthorblockN{Thomas Wiegart, Tobias Prinz, Fabian Steiner, Peihong Yuan}
\IEEEauthorblockA{Institute for Communications Engineering, Technical University of Munich, Germany\\Email: \{thomas.wiegart, tobias.prinz, fabian.steiner, peihong.yuan\}@tum.de}}

\markboth{}{}%
\maketitle

\begin{abstract}
Polar codes are designed for parallel \ac{BiAWGN} channels with an average power constraint. The two main design choices are: the mapping between codeword bits and channels of different quality, and the power allocation under the average power constraint. Information theory suggests to allocate power such that the sum of \ac{MI} terms is maximized. However, a power allocation specific to polar codes shows significant gains.
\end{abstract}


\section{Introduction}
Polar codes were introduced in \cite{St02, Ar09}. They are the first class of codes that achieve the capacity of binary input discrete memoryless channels with a deterministic construction \cite{Ar09}.
 channel.
In \cite{AlTe16} it was shown that the effect of polarization also takes place for non-stationary channels. In this paper, we consider parallel \ac{BiAWGN} channels with an average power constraint \cite[Section 9.4]{cover_elements_it}. Parallel channels naturally arise for \ac{OFDM} transceivers, where a time-frequency resource block has multiple channels of different quality. The model also describes block-fading channels. Polar codes are a natural choice for parallel channels because the different channels can be interpreted as being pre-polarized. 

To develop a basic understanding, we consider the special case of two parallel \ac{BiAWGN} channels.
We address the following two questions: 1) How should the codeword bits be mapped to channels of different quality --- or equivalently, how should one design an interleaver between the codeword bits and the channel. This  has been partially addressed in the literature, e.g., \cite{MaKhLeKa13, LiHoVi17}. Both papers propose a \emph{sorted mapping} that combines two different channels such that each $2 \times 2$ kernel gets one instance of both channels. We also use this mapping, but we show that it does not necessarily minimize the \ac{FER}. 

2) How should power be allocated for a good finite length performance?
 To the best of our knowledge, this has not been considered in the literature yet. We show that the information-theoretic approach of maximizing the achievable rate, also known as mercury/waterfilling \cite{LoTuVe06}, is suboptimal in terms of \ac{FER} for finite length polar codes.  

This work is structured as follows: in Sec.~\ref{sec:preliminaries} we state the system model and preliminaries. In Sec.~\ref{sec:parallel} we discuss the problem of designing polar codes for parallel channels. We provide numerical examples in Sec.~\ref{sec:results} and conclude in Sec.~\ref{sec:conclusion}.

\section{Preliminaries} \label{sec:preliminaries}

 \subsection{Notation}
 We denote random variables by capital letters (e.g., $X$) and deterministic variables or realizations by small letters (e.g., $x$). Deterministic vectors are denoted by a bold italic font with small letters (e.g., $\bm{x}$), while we use a bold italic font with capital letters \mbox{(e.g., $\bm{A}$)} for deterministic matrices and random vectors. We write $\bm{x}_i^j = [x_i, \dots, x_j]$.
\subsection{System Model}
Consider $L$ parallel \ac{BiAWGN} channels
\begin{align}
\label{eq:systemmodel}
Y_i = h_i X_i + N_i, \quad i \in \{1,\dots,L\}
\end{align}
where $Y_i$, $h_i$, $X_i$, and $N_i$ denote the receive signal, the channel coefficient, the transmit signal, and additive white Gaussian noise with $N_i \sim \mathcal{N}(0,1)$, respectively. 
For simplicity, we consider only $L=2$ parallel channels $W_1\!\!: \pp[Y|X][y|x;h_1]$ and $W_2\!\!: \pp[Y|X][y|x;h_2]$. We assume that the channel coefficients $h_i$ are known to the encoder and decoder. The input signals $X_i$ are scaled BPSK symbols, i.e., we have
\begin{align}
    X_i = \sqrt{p_i} S_i, \quad S_i \in \{-1, +1\}.
\end{align}
The value $p_i$ is the power of the transmit signal $X_i$. We consider a common power constraint (see \cite[Section 9.4]{cover_elements_it})
\begin{align}
\label{eq:powerconstr}
\frac{1}{2} (p_1 + p_2) \leq \Pavg .
\end{align}
We combine $\n/2$ uses of each of the two channels to a block of $\n$ channel uses.

\subsection{Polar Codes} \label{sec:polarintro}
Polar codes are linear block codes described by three parameters $[\n, \kc, \Ical]$:  the block length $\n = 2^n, n \in \mathcal{N}$, dimension $\kc$, and a set of information bits $\Ical$ with $\vert \Ical \vert = \kc$. The code rate is $R = \kc / \n$. The input $\bm{u} = [u_1, \dots, u_{\n}] \in \mathbb{F}_2^{\n}$ has an information bit at position $i$ if $i \in \Ical$, and zeros at the remaining positions, i.e., $u_i = 0 \text{ if } i \not\in \Ical$. These bits are called frozen. The codeword $\bm{c} \in \mathbb{F}_2^{\n}$ is generated from $\bm{u}$ by
\begin{align}
    \bm{c} = \bm{u G}_n, \quad \text{with }   \bm{G}_n = \bm{G}_2^{\otimes n} \text{ and } \bm{G}_2 = \begin{bmatrix} 1 & 0 \\ 1 & 1\end{bmatrix}.
\end{align}
$\bm{G}_2^{\otimes n}$ denotes the $n$-th Kronecker power of $\bm{G}_2$. 
The codeword is mapped to BPSK transmit symbols $\bm{x}$ which are transmitted over the channel and received as the vector $\bm{y}$.

With \ac{SC} decoding, the information bits $u_i, i \in \mathcal{I}$, are estimated using $\bm{y}$ and the estimates of the previous bits $\bm{\hat{u}}_1^{i-1}$. The frozen bits are decoded to zero, i.e., $\hat{u}_i = 0$ for $i \not\in \mathcal{I}$.
The \ac{MI} terms $\iI{U_i;\bm{Y} | \bm{U}_1^{i-1}}$ specify the maximum transmission rate over virtual channels  with input $U_i$, output $\bm{Y}$, and known $\bm{U}_1^{i-1}$.
These \ac{MI} terms polarize to being either close to one or close to zero for large $\n$ \cite{Ar09}.
Thus, polar codes are often seen as a transformation of $\n$ channel uses into $\n$ virtual channels with MI either close to one or close to zero. The fraction of virtual channels with MI close to one approaches the capacity of the original channel for large $\n$, and thus polar codes are capacity achieving.

The $\n - \kc$ positions in $\bm{u}$ with smallest MI values are frozen. Polar code design consists of finding these positions.
We use density evolution \cite{MoTa09,MoTa09-2} with a Gaussian approximation \cite{BrKrAs04} to estimate the bit reliabilities.
\begin{figure}[tb]
    \centering
    \footnotesize
    \includegraphics{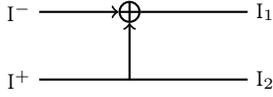} 
    \caption{\ac{MI} of the $2 \times 2$ polar transform.}
    \label{fig:PC2_MI}
\end{figure}
The MI terms can be approximated recursively using the transform depicted in Fig.~\ref{fig:PC2_MI}. The values are given by:
\begin{align}
    \EI^{-} &\approx 1 - \Jfun\left( \sqrt{ \left[\Jfun^{-1}(1-\EI_{1})\right]^2 + \left[ \Jfun^{-1}(1-\EI_{2}) \right]^2 } \right) \\
    \EI^{+} &\approx \Jfun\left( \sqrt{ \left[\Jfun^{-1}(\EI_1)\right]^2 + \left[ \Jfun^{-1}(\EI_2) \right]^2 } \right) 
\end{align}
where the $\Jfun$-function \cite{BrKrAs04} (and its inverse) is approximated numerically \cite{BrRaGr05}. The FER with SC decoding is
\begin{equation} \label{eq:SC-FER}
			\text{FER} = 1 - \prod_{i \in \mathcal{I}} \left( 1 - \operatorname{Pr} \left\{
				\hat{U}_i \neq U_i \given \bm{\hat U}_1^{i-1} = \bm{U}_1^{i-1}
			\right\} \right)
\end{equation}
where $\operatorname{Pr} \left\{
				\hat U_i \neq U_i \given \bm{\hat U}_1^{i-1} = \bm{U}_1^{i-1}
			\right\}$
denotes the probability that the first bit error of a block occurs at bit $i$ (i.e., the probability that the SC decoder makes the wrong decision for bit $i$ given that all previous decisions were correct). We can approximate \eqref{eq:SC-FER} using
\begin{align} \label{SC-FEReinhalb}
    \operatorname{Pr} \left\{
				\hat U_i \neq U_i \given \bm{\hat U}_1^{i-1} = \bm{U}_1^{i-1}
			\right\} \approx  
			\operatorname{Pr}\left\{ \hat U_i \neq U_i \given \bm{U}_1^{i-1}  \right\}
\end{align}
i.e., we assume that a genie-aided decoder was used instead of the real \ac{SC} decoder. Using the MI terms from density evolution, \eqref{SC-FEReinhalb} can be calculated as
\begin{multline} \label{eq:SC-FER2}
			\operatorname{Pr}\left\{ \hat U_i \neq U_i \given \bm{U}_1^{i-1}   \right\}  = 
			\operatorname{Q} \left(
				\frac{1}{2} \Jfun^{-1} \left( \iI{U_i; \bm{Y} | \bm{U}_1^{i-1}} \right)
			\right)
\end{multline}
where $Q(x) = 1/{\sqrt{2 \pi}} \int_x^{\infty} \exp (-{u^2}/{2} ) \mathop{}\!\mathrm{d}u$ denotes the tail distribution function of the normal distribution. The functions in \eqref{eq:SC-FER2} can be approximated numerically.

\subsection{Mercury/Waterfilling} \label{sec:mercury}
Information theory suggests to allocate power such that the achievable rate is maximized, i.e., 
\begin{align}
    \max_{p_1, p_2 \geq 0} \,\, \iI{X_1;Y_1} + \iI{X_2;Y_2} \quad \text{s.t.} \quad \frac{1}{2}(p_1 + p_2) \leq \Pavg .
\end{align}
This optimization problem was solved in \cite{LoTuVe06} for discrete channel input symbols in a (semi-)closed form, and is known as \emph{mercury/waterfilling}. The naming is in analogy to the waterfilling solution for Gaussian inputs \cite[Section 9.4]{cover_elements_it}.
\begin{figure}[tb]
    \centering
    \footnotesize
    \includegraphics{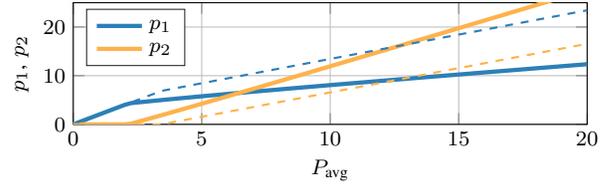} 
    \vspace{-0.2cm}

    \caption{Power allocation for mercury/waterfilling and two parallel \ac{BiAWGN} channels with $h_1 = 0.66$ and $h_2 = 0.33$. For comparison, the waterfilling solution for Gaussian inputs is shown by dashed curves.}
    \label{fig:mercury}
\end{figure}
Fig.~\ref{fig:mercury} shows the mercury/waterfilling solution for two parallel \ac{BiAWGN} channels with channel coefficients $h_1 = 0.66$ and $h_2 = 0.33$. In the low-power regime, the power is allocated only to the better channel. When this channel's \ac{MI} starts to saturate, power is also assigned to the worse channel. For comparison, the waterfilling solution for Gaussian channel inputs is depicted by dashed curves.

\subsection{Normal Approximation}
To take finite length effects into account, we resort to the \ac{NA}~(e.g., \cite[Sec.~II-F]{PoPoVe10}), which is an approximation of the maximum achievable rate for a finite block length $\n$ and reads as
\begin{equation}
R_\text{NA} = C - \sqrt{\frac{V}{\n}} Q^{-1}(\text{FER}) + \frac{1}{2\n}\log_2(\n)
\end{equation}
where $C$ is the capacity of the respective channel and $V$ is the dispersion. The dispersion is defined as $\Var{i(X;Y)}$ with $i(X;Y)$ being the information density. For the considered example of two parallel \ac{BiAWGN} channels we have
\begin{equation}
i(x_1x_2;y_1y_2) = 1 - \sum_{i=1}^{2} \frac{1}{2} \log_2\left(1+\te^{-h_i\sqrt{p_i}s_i y_i}\right).
\end{equation}

\section{Polar Code Design for Parallel Channels} \label{sec:parallel}

\subsection{Problem Statement}
We design polar codes for two parallel \ac{BiAWGN} channels. Each channel is used $\n/2$ times and a polar code of block length $\n$ (which we assume to be a power of $2$) is applied jointly over all channel uses. 
The objective is to minimize the \ac{FER} of a polar code under \ac{SC} decoding.

We optimize the mapping of code word bits to different channels, the set of frozen bits, and the power allocation for $p_1$ and $p_2$ given the average power constraint $P_\text{avg}$. The \ac{FER} under \ac{SC} decoding can be estimated using \eqref{eq:SC-FER} and \eqref{eq:SC-FER2}, such that no Monte-Carlo simulations are necessary.

\subsection{Channel Mappings} \label{sec:mappings}
\begin{figure*}[t]
    \centering
    \subfloat[Sorted mapping]{
        \includegraphics{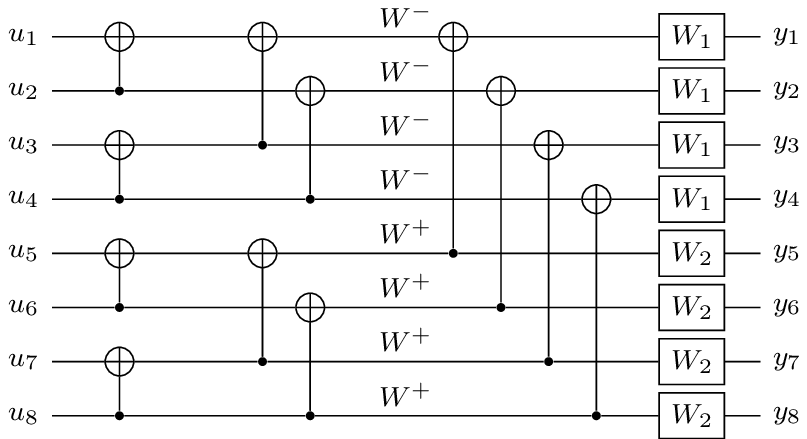} 
        \label{fig:PC8_sorted}
    }
    \hfill
    \subfloat[Alternating mapping]{
        \includegraphics{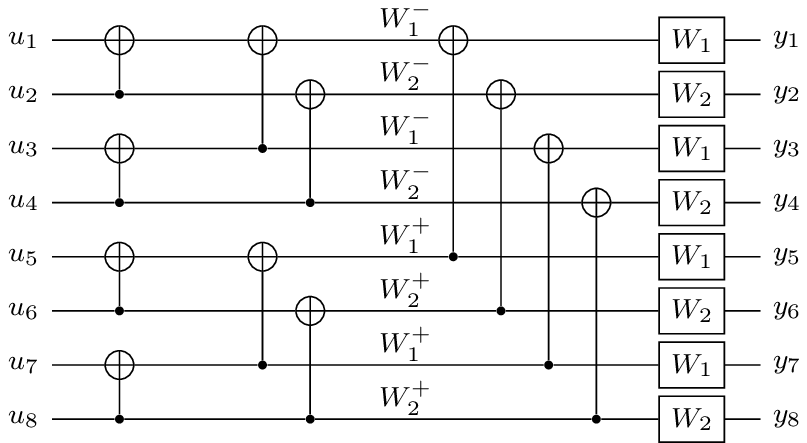} 
        \label{fig:PC8_alternating}
    }

    \caption{Polar codes of length $\n = 8$ over two parallel channels for the sorted mapping and the alternating mapping.}
    \label{fig:PC8}
\end{figure*}
\begin{figure}[tb]
		\centering
		\footnotesize
		\includegraphics{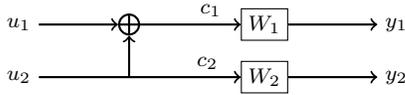} 
		\vspace{-0.1cm}
		\caption{Polar kernel for two parallel channels.}
		\label{fig:parallel:PC2}
	\end{figure}

The mapping of codeword bits to channels has been discussed in \cite{MaKhLeKa13} and \cite{LiHoVi17}. In \cite{MaKhLeKa13}, the authors propose to combine two different channels so that each $2 \times 2$ kernel of the polar code gets one instance of the channel $W_1$ and one instance of the channel $W_2$ (see Fig.~\ref{fig:parallel:PC2} for the $2 \times 2$ kernel and Fig.~\ref{fig:PC8_sorted} for an example of a polar code of length $\n=8$). We denote this mapping as a \emph{sorted} mapping. The other extreme is a mapping we call an \emph{alternating} mapping\footnote{Our nomenclature refers to a non bit-reversal representation of the polar code. In a bit-reversal representation, these two mappings change their roles.}. This mapping combines identical channels as long as possible, i.e., during the first $n-1$ polarization levels (from the channel perspective) for two different channels. An example of this mapping for a polar code of length $\n=8$ is depicted in Fig.~\ref{fig:PC8_alternating}.

	The authors of \cite{LiHoVi17} give reasons for using the sorted mapping. They minimize a bound on the \ac{FER} (similar to \eqref{eq:SC-FER}) with respect to the mapping $\rho$:
	\begin{equation} \label{eq:mappingopt}
		\min_{\rho, \mathcal{I}} \sum_{i \in \mathcal{I}} Z_{n}^{(i)}
	\end{equation}
	where $Z_{n}^{(i)}$ denotes the Bhattacharyya-parameter of the $i$-th virtual channel after $n$ levels of polarization. As solving \eqref{eq:mappingopt} is not feasible, they resort to solving
	\begin{equation}
		\min_{\rho} \sum_{i = 2,4\dots,\n} Z_{1}^{(i)}
	\end{equation}
	i.e., they minimize the sum of even-indexed Bhattacharyya-parameters after the first polarization level. The authors of \cite{LiHoVi17} argue by numerical simulations that this heuristic leads to good results.
	The solution to this relaxed optimization problem is the sorted mapping. However, we figured out that in some scenarios (especially for very short blocks, e.g., for $\n = 8$) the alternating mapping achieves a lower \ac{FER} than the sorted mapping. Thus the sorted mapping is not globally optimal. Nevertheless, we use the sorted mapping for the following reasons:
\begin{figure}[tb]
     \centering
     \footnotesize
    \includegraphics{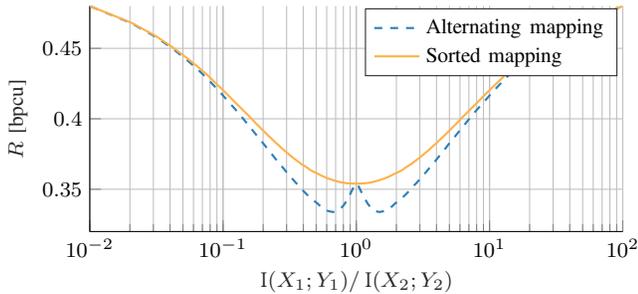} 
    \vspace{-0.2cm}
    \caption{Achievable code rate with SC decoding at a \ac{FER} of $10^{-4}$ for a polar code with block length $\n = \num{16384}$ over two parallel \ac{BiAWGN} channels with average \ac{MI} $1/2 (\iI{X_1;Y_1}  + \iI{X_2;Y_2} ) = 0.5$.}
    \label{fig:rates}
    \vspace{-0.3cm}
\end{figure}
	\begin{itemize}
	    \item After the first level of polarization (from the channel perspective), one obtains two different virtual channels $W^{+}$ and $W^{-}$, see Fig.~\ref{fig:PC8_sorted}. Thus, after the first level, the code behaves like a ``regular'' polar code that also creates two different virtual channels after the first level. This is in contrast to the alternating mapping, where after the first level of polarization there are four different virtual channels, see Fig.~\ref{fig:PC8_alternating}.
	    This insight gives an intuition on how to extend the system to more than two parallel channels, namely by aiming for a ``regular'' polar code after as few levels as possible.
	    \item  Compared to a polar code over identical channels with 
	    \ac{MI} $1/2(\iI{X_1;Y_1} + \iI{X_2;Y_2})$ the code over two parallel channels always leads to stronger polarization in the sense that after the first level of polarization, the virtual channel $W^{-}$ has worse quality than the channel $\bar{W}^{-}$ that would arise from identical channels, and the virtual channel $W^{+}$ has better quality then the channel $\bar{W}^{+}$ that would arise from identical channels. This is shown in Fig.~\ref{fig:rates} where the two mappings are compared in terms of achievable code rate at a fixed \ac{FER} for different channels of constant average MI. When the MI of one channel increases (and thus the MI of the other channel decreases by the same amount), the achievable rate with the sorted mapping increases (for sufficiently large $\n$), whereas the achievable rate with the alternating mapping decreases at first.
	\end{itemize}

\subsection{Frozen Bit Selection} \label{sec:frozenselection}
Suppose the power allocation is fixed, i.e., $p_1$ and $p_2$ are known. 
We use density evolution with Gaussian approximation to select the frozen bits as described in Sec.~\ref{sec:polarintro}. We propagate the MI of the channels through the graphs depicted in Fig.~\ref{fig:PC8}. 

\subsection{Power Allocation}\label{sec:power_allocation}
Next we consider the allocation of powers $p_1$ and $p_2$. From an information theoretic perspective, the powers should be allocated such that the achievable rate (i.e., \ac{MI}) is maximized. This is described in Sec.~\ref{sec:mercury} and the solution is called mercury/waterfilling.

\begin{figure}[bt]
    \centering
    \footnotesize
    \includegraphics{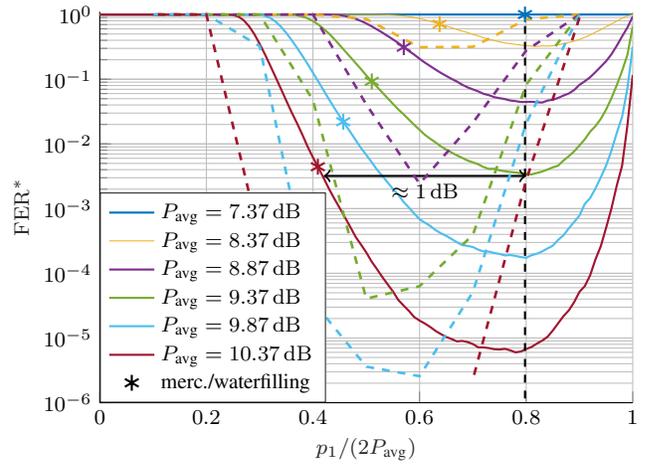} 
    \vspace{-0.3cm}
    \caption{Solid lines depict the \ac{FER} (estimated using \eqref{eq:SC-FER}) versus power allocation for a polar code ($\n = \num{16384}$, $R = 0.5$) over two parallel \ac{BiAWGN} channels with $h_1 = 0.9$ and $h_2 = 0.1$. Dashed lines depict the FER of a 5G \ac{LDPC} code (simulated with a grid size of $0.1$).}
    \label{fig:power1}
\end{figure}

However, it turns out that mercury/waterfilling is not best for finite blocklength polar codes over parallel channels. In particular, we are interested in the power allocation that minimizes the FER of a polar code with fixed parameters (length, dimension, and average power constraint):
\begin{align} \label{eq:FERopt}
    \min_{p_1, p_2 \geq 0} \,\, \FER^{*} (p_1, p_2) \quad \text{s.t.} \quad \frac{1}{2}(p_1 + p_2) \leq P_\text{avg}
\end{align}
where $\FER^{*} (p_1, p_2)$ denotes the \ac{FER} (calculated using \eqref{eq:SC-FER} and \eqref{eq:SC-FER2}) of the polar code with frozen bit indices optimized for the power allocations $p_1$ and $p_2$. 
We assume that the power constraint is fulfilled with equality. Thus, the optimization problem can be re-written as a one dimensional optimization problem in $p_1$, i.e., we have
\begin{align}
    \min_{p_1} \,\, \FER^{*} (p_1, 2 P_\text{avg} - p_1) \quad \text{s.t.} \quad 0 \leq p_1 \leq 2 P_\text{avg} .
\end{align}
The optimization problem can be solved using a simple grid search. 
Fig.~\ref{fig:power1} shows an example of the objective for two parallel channels with channel coefficients $h_1 = 0.9$ and $h_2 = 0.1$. The FER is plotted versus the power allocation $p_1$ (normalized by $2 P_\text{avg}$). Different curves correspond to different power constraints\footnote{The notation of average power in $\SI{}{dB}$ refers to a power gain with respect to the noise random variable with variance $1$, i.e., we calculate $10\log_{10}(P_\text{avg})$.}. 
The power allocations that are given by mercury/waterfilling are depicted by asterisks. The dashed vertical line corresponds to the power allocation given by mercury/waterfilling in the Shannon limit, i.e., the point where $1/2 (\iI{X_1;Y_1} + \iI{X_2;Y_2}) = R$ (in the depicted scenario, the Shannon limit is at $\SI{7.37}{dB}$). As one can see, the FER optimal power allocation is far from the power allocation given by mercury/waterfilling. The difference is several orders of magnitude in \ac{FER}, or more than $\SI{1}{dB}$. The polar-optimal power allocation pushes the good channel further into saturation, i.e., we obtain channels with a stronger pre-polarization. These effects also occur at very long block lengths. Combining polar codes with CRC-aided \ac{SCL} decoding \cite{TaVa15} also leads to similar effects. However, as the \ac{FER} for \ac{SCL} has to be obtained using Monte-Carlo simulations, the optimization is much more complex and we thus focus on optimizing the power allocation for SC decoding.

These results raise the question whether the effects are specific to polar codes or if they originate from a finite number of channel uses. To answer the question, we first compare with an \ac{LDPC} code from the 5G \ac{eMBB} standard~\cite{3gpp-ts-38.212-v15.0.0-17-12a}. The code is derived from basegraph one of the respective standard and has a blocklength of $N = \num{16200}$ and rate $R = 1/2$. As shown in Fig.~\ref{fig:power1} by dashed lines, the optimal power allocation closely follows the assignment given by mercury/waterfilling.

\begin{figure}[tb]
    \centering
    \footnotesize
    \includegraphics{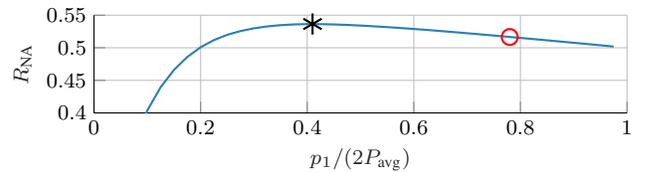} 
    \vspace{-0.3cm}
    \caption{Achievable rate according to normal approximation at a frame error rate of $10^{-4}$ for the scenario from Fig.~\ref{fig:power1} with $P_\text{avg} = \SI{10.37}{dB}$. The power allocation with merucry/waterfilling is denoted by the black asterisk and the polar-optimal power allocation by the red circle.}
    \label{fig:na}
\end{figure}

Secondly, we follow the approach of \cite{PaPa12} and use a finite length bound for power allocation. Fig.~\ref{fig:na} shows the achievable rate according to the normal approximation \cite{PoPoVe10} for the scenario from Fig.~\ref{fig:power1}.
\begin{figure}[tb]
    \centering
    \footnotesize
    \includegraphics{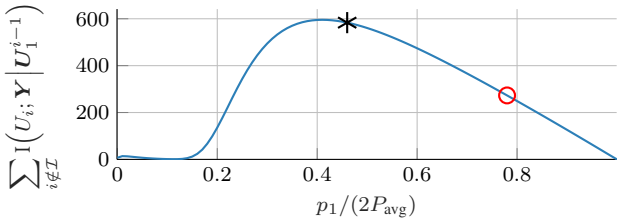} 
    \vspace{-0.3cm}
    \caption{Sum of \ac{MI} terms of frozen bits (choice of frozen bits optimized with Gaussian approximation) for the scenario from Fig.~\ref{fig:power1} and $P_\text{avg} = \SI{10.37}{dB}$.}
    \label{fig:rate_frozen}
\end{figure}
The polar-optimal power allocation (red circle) reduces the achievable rate according to the normal approximation as compared to the mercury/waterfilling solution (black asterisk). Furthermore, the mercury/waterfilling solution is close to the maximum.

From these observations, we conjecture that the effects are inherently linked to polar codes. The behaviour may be partly explained by the following: if bits are frozen whose MI is not zero, then their MI is ``lost'' with SC decoding, as these bits can not be used for information transmission. On the other hand, bits with a MI not close to one need to be frozen to reach a feasible \ac{FER}. Fig.~\ref{fig:rate_frozen} depicts this rate loss for the scenario from Fig.~\ref{fig:power1} with $P_\text{avg} = \SI{10.37}{dB}$. The rate loss with the polar optimal power allocation (red circle) is less than half of the rate loss with mercury/waterfilling (black asterisk). Thus, the polar-optimal power allocation is a tradeoff between rate loss (in terms of achievable rate) by sub-optimal power allocation and rate loss by imperfect polarization. Instead of minimizing the frame error rate one could also maximize the achievable rate of the unfrozen bits, i.e., the rate
\begin{align}
    \nonumber
    \max_{p_1, p_2, \Ical} \,\, \sum_{i \in \Ical} \iI{U_i; \bm{Y} | \bm{U}_1^{i-1}} \quad \text{s.t.} \quad & \frac{1}{2}(p_1 + p_2) \leq \Pavg, \\
    & \vert\Ical\vert = k.
\end{align}
This leads to almost the same results as optimizing the FER \eqref{eq:FERopt}, and brings the power allocation for polar codes back into an information theoretic framework.

\section{Numerical Results} \label{sec:results}
\begin{figure}[tb]
    \centering
    \footnotesize
    \includegraphics{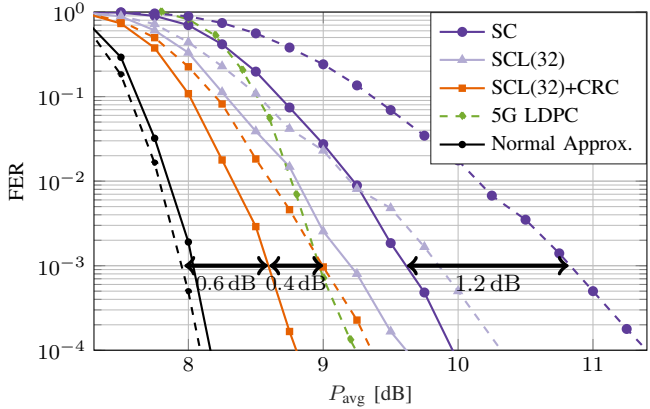} 
    \vspace{-0.3cm}
    \caption{Performance comparison of polar optimal power allocation (solid curves) versus mercury/waterfilling (dashed curves) for a scenario with $h_1 = 0.9$, $h_2 = 0.1$, $\n = \num{16384}$, $R = 0.5$ with SC, SCL, and CRC-aided SCL decoding. For comparison the 5G LDPC code  described Sec.~\ref{sec:power_allocation} and the normal approximation \cite{PoPoVe10} are shown.}
    \label{fig:comparison16k}
\end{figure}
We investigate an extreme case of two parallel channels with $h_1 = 0.9$, $h_2 = 0.1$ and \ac{BPSK}.
The simulation results are depicted in Fig.~\ref{fig:comparison16k}. A polar code of block length $\n = \num{16384}$ is used. The figure shows the FER versus the average power. With \ac{SC} decoding, the polar code with optimized power allocation outperforms the polar code with mercury/waterfilling by $\SI{1.2}{dB}$ at a FER of $10^{-3}$. For \ac{SCL} decoding \cite{TaVa15} with list size $L=32$, the qualitative behaviour stays the same, but the gap between the two power allocations shrinks to approximately $\SI{0.7}{dB}$. The SC decoded polar code with optimized power allocation outperforms the SCL decoded polar code with mercury/waterfilling. When combining SCL decoding with an outer CRC with $\SI{20}{bits}$, the polar code with power allocation optimized for SC decoding still outperforms the polar code with mercury/waterfilling by $\SI{0.4}{dB}$. It outperforms the 5G LDPC code by about $\SI{0.4}{dB}$ and operates approximately  $\SI{0.6}{dB}$ away from the normal approximation \cite{PoPoVe10}.

\vspace{-0.05cm}
\section{Conclusion}
\label{sec:conclusion}
We proposed a novel approach to allocate power for polar codes over parallel channels with an average power constraint. We showed significant gains in terms of FER as compared to power allocation by mercury/waterfilling. We elaborated on the design of polar codes for parallel channels and the mapping between codeword bits and channels of different quality. Future work involves a study of more than two parallel channels, including the design of the mapping between codeword bits and channels. A further research topic is the power allocation for polar codes with higher order modulation.

\section*{Acknowledgement}
The authors would like to thank Dr.~Gianluigi Liva for helpful and enlightning discussions regarding the error probability approximations in \eqref{SC-FEReinhalb} and \eqref{eq:SC-FER2}.

\bibliographystyle{IEEEtran}
\bibliography{IEEEabrv,confs-jrnls,literature}

\begin{thebibliography}{10}
\providecommand{\url}[1]{#1}
\csname url@samestyle\endcsname
\providecommand{\newblock}{\relax}
\providecommand{\bibinfo}[2]{#2}
\providecommand{\BIBentrySTDinterwordspacing}{\spaceskip=0pt\relax}
\providecommand{\BIBentryALTinterwordstretchfactor}{4}
\providecommand{\BIBentryALTinterwordspacing}{\spaceskip=\fontdimen2\font plus
\BIBentryALTinterwordstretchfactor\fontdimen3\font minus
  \fontdimen4\font\relax}
\providecommand{\BIBforeignlanguage}[2]{{%
\expandafter\ifx\csname l@#1\endcsname\relax
\typeout{** WARNING: IEEEtran.bst: No hyphenation pattern has been}%
\typeout{** loaded for the language `#1'. Using the pattern for}%
\typeout{** the default language instead.}%
\else
\language=\csname l@#1\endcsname
\fi
#2}}
\providecommand{\BIBdecl}{\relax}
\BIBdecl

\bibitem{St02}
\BIBentryALTinterwordspacing
N.~Stolte, ``Rekursive codes mit der {P}lotkin-konstruktion und ihre
  decodierung,'' Ph.D. dissertation, Technische Universit{\"a}t, Darmstadt,
  Januar 2002. [Online]. Available:
  \url{http://tuprints.ulb.tu-darmstadt.de/183/}
\BIBentrySTDinterwordspacing

\bibitem{Ar09}
E.~Ar{\i{}}kan, ``Channel polarization: a method for constructing
  capacity-achieving codes for symmetric binary-input memoryless channels,''
  \emph{{IEEE} Trans. Inf. Theory}, vol.~55, no.~7, pp. 3051--3073, July 2009.

\bibitem{AlTe16}
M.~Alsan and E.~Telatar, ``A simple proof of polarization and polarization for
  non-stationary memoryless channels,'' \emph{{IEEE} Trans. Inf. Theory},
  vol.~62, no.~9, pp. 4873--4878, Sept 2016.

\bibitem{cover_elements_it}
T.~M. Cover and J.~A. Thomas, \emph{Elements of {{Information Theory}}},
  2nd~ed.\hskip 1em plus 0.5em minus 0.4em\relax {John Wiley \& Sons, Inc.},
  2006.

\bibitem{MaKhLeKa13}
H.~Mahdavifar, M.~El-Khamy, J.~Lee, and I.~Kang, ``Compound polar codes,'' in
  \emph{Inf.~Theory and Appl.~Workshop}, Feb 2013, pp. 1--6.

\bibitem{LiHoVi17}
S.~Liu, Y.~Hong, and E.~Viterbo, ``Polar codes for block fading channels,'' in
  \emph{IEEE Wireless Commun. and Netw. Conf. Workshops}, March 2017, pp. 1--6.

\bibitem{LoTuVe06}
A.~Lozano, A.~M. Tulino, and S.~Verdú, ``Optimum power allocation for parallel
  {G}aussian channels with arbitrary input distributions,'' \emph{{IEEE} Trans.
  Inf. Theory}, vol.~52, no.~7, pp. 3033--3051, July 2006.

\bibitem{MoTa09}
R.~Mori and T.~Tanaka, ``Performance and construction of polar codes on
  symmetric binary-input memoryless channels,'' in \emph{IEEE Int.\ Symp.\
  Inf.\ Theory (ISIT)}, June 2009, pp. 1496--1500.

\bibitem{MoTa09-2}
------, ``Performance of polar codes with the construction using density
  evolution,'' \emph{{IEEE} Commun. Lett.}, vol.~13, no.~7, pp. 519--521, July
  2009.

\bibitem{BrKrAs04}
S.~ten Brink, G.~Kramer, and A.~Ashikhmin, ``Design of low-density parity-check
  codes for modulation and detection,'' \emph{{IEEE} Trans. Commun.}, vol.~52,
  no.~4, pp. 670--678, April 2004.

\bibitem{BrRaGr05}
F.~Brannström, L.~K. Rasmussen, and A.~J. Grant, ``Convergence analysis and
  optimal scheduling for multiple concatenated codes,'' \emph{{IEEE} Trans.
  Inf. Theory}, vol.~51, no.~9, pp. 3354--3364, Sept 2005.

\bibitem{PoPoVe10}
Y.~Polyanskiy, H.~V. Poor, and S.~Verdú, ``Channel coding rate in the finite
  blocklength regime,'' \emph{{IEEE} Trans. Inf. Theory}, vol.~56, no.~5, pp.
  2307--2359, May 2010.

\bibitem{TaVa15}
I.~Tal and A.~Vardy, ``List decoding of polar codes,'' \emph{{IEEE} Trans. Inf.
  Theory}, vol.~61, no.~5, pp. 2213--2226, May 2015.

\bibitem{3gpp-ts-38.212-v15.0.0-17-12a}
``{{3GPP TS}} 38.212 {{V15}}.0.0: {{Multiplexing}} and channel coding,'' Dec.
  2017.

\bibitem{PaPa12}
J.~Park and D.~Park, ``A new power allocation method for parallel {AWGN}
  channels in the finite block length regime,'' \emph{{IEEE} Commun. Lett.},
  vol.~16, no.~9, pp. 1392--1395, September 2012.

\end{thebibliography}

\end{document}